\documentclass[12pt,a4 paper, one side]{article} 
\usepackage{tikz}
\usepackage{amsmath,amssymb,latexsym}
\usepackage{xcolor}
\usepackage{hyperref}
\usepackage{graphicx}
\usepackage{epstopdf}
\usepackage{enumerate}
\usepackage[utf8]{inputenc}
\usepackage[T1]{fontenc}
\usepackage{amsmath}
\usepackage[leftcaption]{sidecap}
\usepackage{amssymb}
\usepackage{graphicx}
\usepackage{booktabs,dcolumn,caption}
\usepackage{rotating}
\usepackage{amsthm}
\usepackage{tikz}
\usepackage{caption}
\usepackage{subcaption}
\usepackage[most]{tcolorbox}
\begin{document}
	\date{}
	\title{Astrophysical implications of an eternal homogeneous gravitational collapse model with a parametrization of expansion scalar}
	\maketitle
	\begin{center}
		\author{ Annu Jaiswal\footnote{annujais3012@gmail.com}, Rajesh Kumar\footnote{rkmath09@gmail.com}, Sudhir Kumar Srivastava\footnote{sudhirpr66@rediffmail.com}
			\\
			Department of Mathematics and Statistics,\\
			Deen Dayal Upadhyaya Gorakhpur University, Gorakhpur, INDIA.\\
			
			S.K.J.Pacif\footnote{shibesh.math@gmail.com}
			\\Centre for Cosmology and Science Popularization (CCSP), SGT University, Delhi-NCR, Gurugram 122505, Haryana, INDIA} 
	\end{center}

	\begin{abstract}
		In this study, we continue our previous work (Annu et al., 2023) by introducing a novel parametrization of the expansion scalar $\Theta $ as a rational function of time $t$. The paper provides a comprehensive analysis of a homogeneous gravitational collapsing system, wherein the exact solutions of the Einstein field equations (EFEs) are determined using a new parametrization of $\Theta$ in a model-independent way. The model is especially significant for the astrophysical applications because we have addressed the physical and geometrical quantities of the model in terms of Schwarzschild mass $M$. We have estimated the numerical value of the model parameter involved in the functional form of $\Theta$-parametrization using the masses \& radii data of some massive stars namely, Westerhout 49-2, BAT99-98, R136a1, R136a2, WR 24, Pismis 24-1, $\lambda$ Cephei, $\alpha$  Camelopardalis, $\beta$ Canis Majoris. We have presented theoretical investigations about such astrophysical stellar systems. The formation of an apparent horizon is also studied for the collapsing system, and it has been shown that  our model produces a continuing collapsing scenario of star (an eternal collapsing object).
		
	\end{abstract}
	
	
	\section{ Introduction}\label{sec1}
	Gravitational collapse, specifically the end state of a massive collapsing star is one of the challenging issues for theoretical astrophysicists. Most of the known astrophysical objects such as stars, white dwarfs and neutron stars, emerge from the gravitational collapse of stellar systems. It is generally speculate that the collapsing stars with ultimate masses of $5M_{\odot }$  or greater should evolve into black holes \cite{RBB20} but when sufficiently massive stars exhaust their nuclear fuels what would be its end state, is a critical problem in astronomy and astrophysics for decades. To discuss the possible end state of such a continued gravitational collapse, one must study dynamical collapse scenarios of star within the framework of a gravitation theory such as Einstein’s theory of general relativity, etc.  During  nuclear fusion process in massive stars, the gravitational collapse cannot be countered by any external heat pressure, and the star starts collapse to a space-time singularity \cite{SW73}-\cite{RP69}. According to Penrose's cosmic censorship conjecture (CCC), the singularity formed by gravitational collapse should be hidden behind the horizon, implying that BH is the only feasible end state of collapse \cite{RP69}. Further, since the CCC has no proper mathematical proof and there are also various models related to the gravitational collapse of matter have been also constructed so far where one encounters a naked singularity (NS) \cite{PS2007}. However, it is also believed that no past extendible non-spacelike geodesic can exist between the singularity and any point on the space-time manifold i.e., no non-spacelike geodesic could have a positive tangent at the singularity- strong cosmic censorship hypothesis \cite{PS2007}.
	
	\par
	Thus, now a days the final stage of a collapsing massive star  remains unsolved problems in astrophysics. It is important to mention here  that the Oppenheimer and Snyder (OS) model \cite{JR39} serves as the framework for the notion in the inevitable development of BH as far as solutions are concerned. OS initiated the study of gravitational collapse with an FLRW like metric and later on several authors extended this study of gravitational collapse (\cite{HD10}-\cite{RS18} and many more). Since the motion of collapsing fluids in general relativity is determined by a number of variables, including shear tensor, vorticity tensor (which vanishes in present case),  acceleration vector and  expansion scalar ($\Theta $). The dynamics of the stellar system are determined by these kinematical quantities that evolve throughout the gravitational collapse. Recently, authors \cite{RJ22}-\cite{RJ23a} have studied a new class of gravitational collapse with uniform expansion scalar, which may describes the interesting scenario of collapsing stellar systems and may also have many astrophysical consequences. The formulation of Einstein's field equations (EFEs) has allowed theoretical physicists to suggest various models of high-gravity astrophysical phenomena such as quasars, black holes, and other super-dense objects generated by gravitational collapse.
	
	\par
	The current work aims to examine the homogenous collapse of perfect fluid distributions from entirely new approaches and discuss the solution of EFEs by using boundary-conditions. In our previous work \cite{RJ23a}, we have used the exponential and power law parametrization of $\Theta $. Here, we have introduced a new $\Theta$-parameterization following the collapsing configurations and calculate the numerical value of model parameter for the massive star namely, Westerhout 49-2, BAT99-98, R136a1, R136a2, WR 24, Pismis 24-1, $\lambda$ Cephei, $\alpha$  Camelopardalis, $\beta$ Canis Majoris. Further, a comprehensive discussion of an apparent horizon  and nature singularity formation are carried out in this work. The model present a new schenario of collapsing system called-eternal collapse phenomenon.
	
	\par
	The paper is organized as follows- after introduction, in section (\ref{sec2}), we have defined the basic formalisms and EFE for FLRW space-time metric with perfect fluid distributions and the exterior vacuum region of the system is described by the Schwarzschild space-time. Section (\ref{sec3}) introduces the $\Theta$-parametrization of rational function of $t$. In section (\ref{sec4}), we have discussed the exact solutions of EFE and obtain all the parameters in terms of Schwarzschild mass $M$ using boundary condition. We have also presented the solution with the graphical representation for the massive stars of their known masses and radii. The section (\ref{sec5}) includes the discussion of formation of apparent horizon and eternal collapse phenomenon. The last section (\ref{sec6}) contains the discussion and concluding remarks.
	
	
	\section{Formalism of Gravitational Collapse}\label{sec2}
	\subsection{Metric and the basic equations}\label{subsec2.1}
	We consider that the spherically symmetric space time inside the collapsing stellar system (e.g., star) as homogeneous and isotropic FLRW  metric
	\begin{equation}	
		ds_{-}^2 = -dt^2 + a^2(t)dr^2 + R^2(t,r) \left( d\theta^2 + \sin^2\theta d\phi^2 \right)
		\label{eq1}
	\end{equation}	
	where  $a(t)$ is the scale  factor  and  $R(t,r) = r a(t)$ is the geometrical radius of the stellar system. Here the coordinate is taken as  $x^{i} = (t,r,\theta,\phi)$, $i = 0,1,2,3$ and the comoving four velocity vector $V^{i}$ satisfy
	\begin{equation}
		V^{i} V_{i} = -1 ~\quad \mbox{where}~\quad  V_i = (-1, 0, 0, 0)
		\label{eq2}
	\end{equation}
	
	The matter inside the stellar system is considered to be perfect fluid distribution described by the energy-momentum tensor 
	\begin{equation}
		T_{ij} = (p+\rho) V_i V_j + p g_{ij}
		\label{eq3}
	\end{equation}
	where $\rho$ and $p$ are respectively the energy density and pressure of the fluid distribution. 
	\par
	
	Since for collapsing configuration the areal velocity $\frac{\dot{R}}{R} < 0$ and the collapsing rate of the star is described by the expansion-scalar($\Theta$) 
	\begin{equation}
		\Theta = V^i_{;i} = 3\frac{\dot{R}}{R}
		\label{eq5}
	\end{equation}
	where dot $(.)$ denotes the derivative with respect to time $t$. 
	\par
	The Einstein's field equation	
	\begin{equation*}
		R_{ij} - \frac{1}{2}  \mathcal{R} g_{ij}=\mathcal{X} T_{ij}
		\label{eq6}
	\end{equation*}
	for the present case yields the following independent equations (where  $\mathcal{X} = \frac{8 \pi G}{c^4}$ )
	\begin{equation}
		\mathcal{X} \rho = \frac{1}{3} \Theta^2 
		\label{eq7}
	\end{equation}
	
	\begin{equation}
		\mathcal{X} p = -\frac{1}{3} ( \Theta^2 + 2 \dot{\Theta})
		\label{eq8}
	\end{equation}
	The mass function $m(t,r)$ for the spherically symmetric collapsing system at any instant is given by \cite{me70}   \\
	\begin{equation}
		m(t,r) = \frac{1}{2} R \left(1 + R_{,i} R_{,j} g^{ij}\right) = \frac{1}{18} \Theta^2 R^3 
		\label{eq11}
	\end{equation}	
	where the comma (,) denotes the  partial derivative.
	
	\subsection{Junction condition and the Kretschmann Curvature}\label{subsec2.2}
	In general relativity, accoding to the Jebsen-Birkhoff's theorem, the Schwarzschild solution is the  exact solution of vaccum Einstein's field equations decribing the gravitational fields  exterior to  a spherically symmetric star. The Schwarzschild metric is
	\begin{equation}
		ds^{2}_{+} = -\left(1-\frac{2M}{\mathbf{r}}\right) d\tau^{2} + \frac{d\mathbf{r}^{2}}{\left(1-\frac{2M}{\mathbf{r}}\right)} + \mathbf{r}^{2} d\Omega^{2}
		\label{eq13}
	\end{equation}
	where $M$ represent the Newtonian mass of star (also known as Schwarzschild mass) and the coordinate of exterior space-time is $(\tau, \mathbf{r}, \theta, \phi)$.
	\par
	The boundary hyper-surface $\Sigma$ separates the  stellar system into the interior $(ds^{2}_{-})$ and the exterior $(ds^{2}_{+})$ spacetime metric. The matching of interior metric to the exterior Schwarzschild metric on the hyper-surface $\Sigma$ yield the boundary conditions \cite{NO16}
	\begin{equation}
		m\left(t,r\right) \overset{\Sigma}{=} M
		\label{eq14}
	\end{equation}
	The eq.(\ref{eq14}) shows that the mass-function $m(t,r)$ must be equal to the Schwarzschild mass  $M$ on $\Sigma$ i.e., initially at $t=t_0, r=r_0$ the mass of collapsing star is considered $M$.
	\par
	The Kretschmann curvature is a quadratic scalar invariant derived by the full-contraction of Riemann curvature tensor \cite{SO02}
	\begin{equation}
		\mathcal{K} = R_{i j k \delta} R^{i j k \delta}
		\label{eq16}
	\end{equation}
	where $R_{i j k \delta}$ denotes Riemann curvature tensor. For the metric (\ref{eq1}), we have
	\begin{equation}
		\mathcal{K} = \dfrac{12 \left(\dot{a}^4 + a^2 \ddot{a}^2\right)}{a^4}
		\label{eq19}
	\end{equation}  
	The singularity of space-time is identified when the  Kretschmann curvature $\mathcal{K}$ diverges uniformly.
	\section{Parametrization of $\Theta $}\label{sec3} 
	The system of differential equations (\ref{eq7})-(\ref{eq8}) possess only two independent equations with three unknowns $a(t)$, $p(t)$ and $\rho (t)$. Therefore, it requires one more constraints for the complete determination of the solution of EFEs. In fact, a critical analysis of the solution techniques of EFEs in general relativity (or, in modified gravity theories) is the parametrization of geometrical$/$ physical parameters and various schemes of parametrization are used in cosmology (\cite{RJ23a}, \cite{R17}-\cite{R18} and references their in).
	\par
	
\begin{figure}[h]
		\centering
		\includegraphics[scale=0.87]{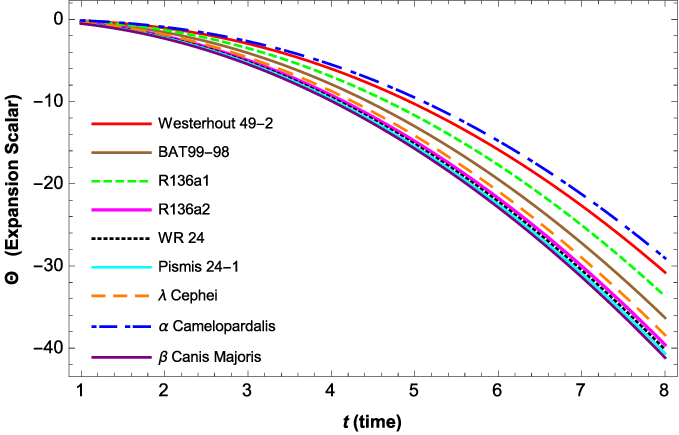}
		\caption{Collapsing configuration: The expansion scalar ($\Theta$) is plotted with respect to time coordinate \textbf{$t$} for the nine massive stars corresponding to values of parameter $\gamma$ given in Table \ref{table1} with following eq.(\ref{eq20}).}
		\label{fig1}
	\end{figure}

In the evolution of stellar system (e.g., star), because of the nuclear fusion process in the core of star, it loses its equilibrium-stage and started to collapse under its own gravity \cite{PS2007}. During the collapsing process, the internal thermal pressure (which arises during nuclear reaction of H or, He molecules) decreases and then the external pressure (which is due to the gravitational mass of star) dominate over it. In this way, the collapsing velocity (collapse rate) of the star increases which tends to draw matter inward toward the centre of gravity. According to GR, in the collapsing system, two kinds of motion (velocities) occur namely $\frac{\dot{R}}{R}$ which, as mentioned before, measures the variation of the radius $R$ per unit proper time and, another $\dot{(\delta l)}$, the variation of the infinitesimal proper radial distance $\delta l$ between two neighbouring fluid particles per unit of proper time. Since, the expansion scalar $\Theta $ is defined as the rate of change of elementary fluid particles which describes collapsing rate of system and the FLRW homogenous gravitational collapse requires the motion of the fluid  particle to be uniform, independent of radial distance $r$. In such a way, if one observes carefully that the collapsing rate increases and hence the expansion scalar $\Theta $ also increases with $t$, where $\Theta =3\frac{\dot{R}}{R}<0$ for collapsing configuration. Therefore, for the parameterization of $\Theta $, one can consider it as a function of $t$, which precisely explains its notion and also depicts the collapsing configuration (fig.\ref{fig1}). Recently, we have considered some schemes of $\Theta$-parametrization to solve the field equations for  the collapsing system \cite{RJ23}-\cite{RJ23a}. In the present work, we have introduced the  $\Theta $-parameterization as a rational function of $t$ as follows,
	
	\begin{equation}
		\Theta = -\frac{16 \gamma  t^3}{15 \left(\frac{8 \gamma  t}{5}+1\right)},~\quad~\gamma \ge 0
		\label{eq20}
	\end{equation}
	where $\gamma$ is a model parameter to be determined by using the observational data of some known massive stars in Table \ref{table1}.
	
	
	\section{Exact solution of Einstein field equation} \label{sec4}
	The eq.(\ref{eq20}) is the additional constraint, has been	used for the solution of EFEs (\ref{eq7})-(\ref{eq8}). By using eq.(\ref{eq5}) into (\ref{eq20}) and integrating, we have 
	
	\begin{equation}
		a(t) = k e^ {-\frac{16}{9} \gamma  \left(\frac{25 t}{512 \gamma ^3}-\frac{5 t^2}{128 \gamma ^2}+\frac{t^3}{24 \gamma } -\frac{125 \log (8 \gamma  t+5)}{4096 \gamma ^4}\right)}
		\label{eq21}
	\end{equation}
	where $k$ is an integrating constant. In order to determine the value of  $k$, we use the boundary condition (\ref{eq14}).\\
	Let us assume the star begins to collapse initially at ($t_{0}, r_{0}$), then  apply the condition (\ref{eq14}) by taking use of  eqs.(\ref{eq11}) and (\ref{eq21}), we have
	\begin{equation}
		\frac{128}{81} \gamma^2  {r_0}^3 {t_0}^6 e^{-\frac{{t_0} \left(64 \gamma^2 {t_0}^2-60 \gamma{t_0}+75\right)}{288 \gamma^2}} (8 \gamma  {t_0}+5)^{-2+\frac{125}{768 \gamma^3}} k^3=M
		\label{eq22}
	\end{equation}
	solving above for $k$, we obtain 
	\begin{equation}
		k =  \frac{3 {\left(\frac{3}{2}\right)^{\frac{1}{3}} M^{\frac{1}{3}} e^{\frac{{t_0} \left(64\gamma^2 {t_0}^2-60 \gamma{t_0}+75\right)}{864 \gamma^2}} (8 \alpha {t_0}+5)^{\frac{1}{3} \left(2-\frac{125}{768 \gamma^3}\right)}}}{4 \gamma^{\frac{2}{3}}{r_0} {t_0}^2}
		\label{eq23}
	\end{equation} 
	Substituting the value of $k$ into  eq.(\ref{eq21}) we obtain
	
	\begin{equation}
		a(t)  = \frac{ {(\frac{81 M}{2})}^{\frac{1}{3}} (8\gamma  {t_0}+5)^{\frac{2}{3}-\frac{125}{2304\gamma^3}}}{4 \gamma^{\frac{2}{3}} {r_0} {t_0}^2} e^{\frac{375 \log (8 \gamma  t+5)-8 \gamma  \left(64 \gamma ^2 t^3-60 \gamma  t^2+75 t+\text{t0} \left(-64 \gamma ^2 \text{t0}^2+60 \gamma  \text{t0}-75\right)\right)}{6912 \gamma ^3}}
		\label{eq24}
	\end{equation}
	Now, by using eqs.(\ref{eq20}) and (\ref{eq24}),  we obtain from eqs.(\ref{eq7})-(\ref{eq8}) that 
	\begin{equation}
		\mathcal{X} p = -\frac{32 \gamma t^2 \left(48 \gamma  t+45- 8 \gamma t^4\right)}{27  (8 \gamma  t+5)^2}
		\label{eq25}
	\end{equation}
	the negative sign in $p$ indicates the pressure towards the center of star in collapsing confuguration.
	\begin{equation}
		\mathcal{X} \rho = \frac{256 \gamma^2 t^6}{27 (8 \gamma t+5)^2}
		\label{eq26}
	\end{equation}
	
	Using eqs.(\ref{eq20}) and (\ref{eq24}) into (\ref{eq11}), we obtain
	\begin{equation}
		m =\frac{M}{r_0^3 t_0^6} (8 \gamma  t+5)^{2-\frac{125}{768 \gamma ^3}} r^3 t^6 (8 \gamma  t+5)^{\frac{125}{768 \gamma ^3}-2} e^{\frac{-64 \gamma ^2 t^3+60 \gamma  t^2-75 t+t_0 \left(64 \gamma ^2 t_0^2-60 \gamma  t_0+75\right)}{288 \gamma ^2}}
		\label{eq27}
	\end{equation}
	Taking partial derivatives with respect to $t$ and $r$ respectively gives the rate of change of mass ($\dot{m}$) and mass-gradient ($m'$) 
	
	\begin{equation}
		\begin{split}
			\dot{m}=\frac{2 M r^3 t^5}{3 {r_0}^3 {t_0}^6} (8 \gamma   t+5)^{-3+\frac{125}{768 \gamma  ^3}} \left(48 \alpha  t+45-8 \gamma   t^4\right) (8 \gamma  {t_0}+5)^{2-\frac{125}{768\gamma ^3}} &\\ e^{\frac{-64 \gamma  ^2 t^3+60 \gamma  t^2-75 t+\text{t0} \left(64 \gamma  ^2 {t_0}^2-60 \gamma   {t_0}+75\right)}{288 \gamma  ^2}}
			\label{eq27a}
		\end{split}
	\end{equation}
	\begin{equation}
		\begin{split}
			m'=	\frac{3 M r^2 t^6}{r_0^3 t_0^6} (8 \gamma   t+5)^{\frac{125}{768 \gamma ^3}-2} (8 \gamma  t_0+5)^{2-\frac{125}{768\gamma  ^3}} e^{\frac{-64\gamma ^2 t^3+60 \gamma   t^2-75 t+t_0 (64 \gamma ^2 t_0^2-60\gamma t_0+75)}{288\gamma  ^2}}
			\label{eq111}
		\end{split}
	\end{equation}
	Also from eq.(\ref{eq19}), we obtained the Kretschmann curvature
	\begin{equation}
		\begin{split}
			\mathcal{K} = \frac{1024 \gamma^2 t^4}{2187 (8 \gamma t+5)^4} ( 512 \gamma^2 t^8 -4608 \gamma^2 t^5- 4320\gamma  t^4 &\\ +20736 \gamma^2 t^2	+ 38880 \gamma t+18225 )
			\label{eq28}
		\end{split}
	\end{equation}
	The collapsing acceleration($\frac{\ddot{a}}{a}$) is obtained from eq. (\ref{eq24})
	\begin{equation}
		\frac{\ddot{a}}{a}=\frac{16  \gamma  t^2 \left(16  \gamma  t^4-144 \gamma  t-135\right)}{81 (8  \gamma t+5)^2}
		\label{eq91}
	\end{equation}
	It can be seen from eqs.(\ref{eq24})-(\ref{eq91}) that all the quantities namely $a$, $\,p$, $\rho $, $m$, $\mathcal{K}$ and $\frac{\ddot{a}}{a}$ are obtained in terms of mass $M$ of star. Therefore, the solution is applicable to the studies of the known stars whose masses and radius are given \cite{ob1}-\cite{ob15}. By knowing of model parameters $\gamma $, one can discuss the dynamics of such known stars and explored towards their astrophysical significances.
\subsection{Estimation of model parameter $\gamma$ for known massive stars}\label{subsec4.1}
We assume that the collapse of star begins initially at ($t_0, r_0$), then surface radius of star become $R_0 = R(t_0, r_0) = r_0 a(t_0)$ and is given by
\begin{equation}
R_0 = \frac{ {(\frac{81 M}{2})}^{\frac{1}{3}} (8\gamma  {t_0}+5)^{\frac{2}{3}-\frac{125}{2304\gamma^3}}}{4 \gamma^{\frac{2}{3}} {t_0}^2} e^{\frac{375 \log (8 \gamma  t+5)-8 \gamma  \left(64 \gamma ^2 t^3-60 \gamma  t^2+75 t+\text{t0} \left(-64 \gamma ^2 \text{t0}^2+60 \gamma  \text{t0}-75\right)\right)}{6912 \gamma ^3}}
\label{eq91a}
\end{equation}
Using the observational data of masses ($M$) and radii ($R_0$) for known massive stars (summarize in  Table \ref{table1}) into eq.(\ref{eq91a}), we obtain
\begin{equation}
1.00023 \frac{(5+8 \gamma)^{\frac{2}{3}}}{\gamma^{\frac{1}{3}}} = 10.222918629 \gamma^{\frac{1}{3}}~\quad~ \mbox{for Westerhout 49-2}
\label{eq91b}
\end{equation}
by assumimg the initial time $t_0 = 1$, Solar mass $M_\odot$ and Solar radius $R_\odot$ to be unit $1$.  Solving eq(\ref{eq91b}) we get $\gamma = 0.202544$ approximately. Similarly, the numerical value of $\gamma$ can be evaluated for other massive stars as shown in Table \ref{table1}.
	\captionsetup{labelsep=newline,
		singlelinecheck=false,
		skip=0.333\baselineskip}
	\newcolumntype{d}[1]{D{.}{.}{#1}} 
	\renewcommand{\ast}{{}^{\textstyle *}} 
	\renewcommand\arraystretch{1.5}
	\begin{table}[h]
		\caption{\textbf{Numerical value of parameter $\gamma$ for  known masses $M$ and radii $R_0$ of the  massive stars} }
		\label {table1}
		\begin{tabular}{p{5.2cm}p{2.9cm}p{2.9cm}p{2cm}}
			\hline\specialrule{1.1pt}{0pt}{0pt}
			\textbf{Massive Star} &  $M (M_\odot)$   &   $ R_0 (R_\odot)$ & $\gamma$  \\
			\hline\specialrule{1.1pt}{0pt}{0pt}
			Westerhout 49-2\cite{ob1}\cite{ob2} &	250	& 55.29 &	0.202544 \\ 
			BAT99-98 \cite{ob3} &	226 &	37.5 &	0.446334\\
			R136a1\cite{ob4}\cite{ob5}	&196&	42.7&	0.293187\\
			R136a2\cite{ob4}\cite{ob6} &	151	&25.2&	1.0119\\
			WR 24 \cite{ob7}\cite{ob8}&	114&	21.73&	1.27352\\
			Pismis 24-1\cite{ob9}\cite{ob10} & 74 & 18 & 1.58239\\
			$\lambda$ Cephei\cite{ob11}\cite{ob12} &	51.4&	19.5&	0.704378\\
			$\alpha$ Camelopardalis\cite{ob12}\cite{ob13} &	37.6&	32.5&	0.166759\\
			$\beta$ Canis Majoris\cite{ob14}\cite{ob15} &	13.5 & 9.7 & 2.1404 \\ \addlinespace 
			\hline\specialrule{1.1pt}{0pt}{0pt} 
		\end{tabular}
	\end{table}
	\begin{figure}[h]
	\centering
	\includegraphics{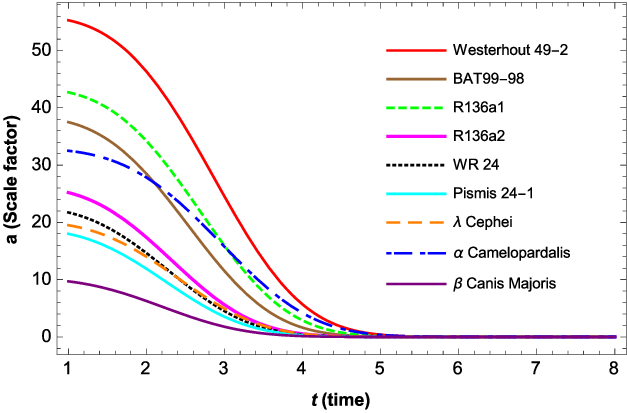}
	\caption{The scalar factor ($a$) is plotted with respect to time coordinate \textbf{$t$} for the nine massive stars corresponding to the values of parameter $\gamma$, masses $M$, initial coordinate $(t_0, r_0) = (1, 1)$ given in Table \ref{table1} with following eq. (\ref{eq24}).}
	\label{fig2}
\end{figure}
\begin{figure}[h]
	\centering
	\includegraphics{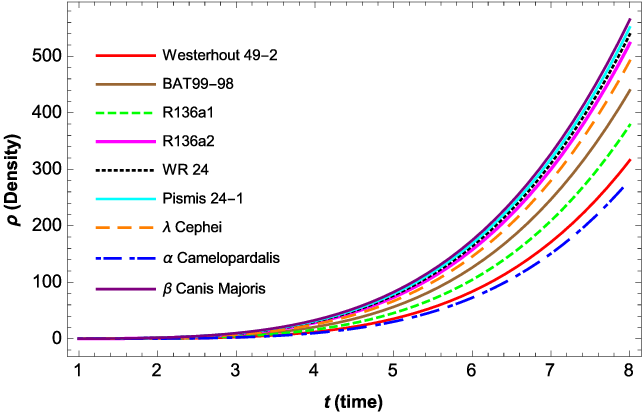}
	\caption{The energy density ($\rho$) is plotted with respect to time coordinate \textbf{$t$} for the nine massive stars corresponding to the values of parameter $\gamma$, masses $M$, initial coordinate $(t_0, r_0) = (1, 1)$ given in Table \ref{table1} with following eq. (\ref{eq26}).}
	\label{fig3}
\end{figure}
\begin{figure}[h]
	\centering
	\includegraphics{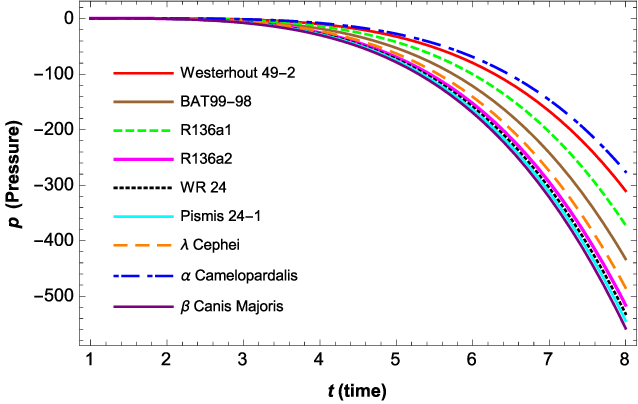}
	\caption{The pressure ($p$) is plotted with respect to time coordinate \textbf{$t$} for the nine massive stars corresponding to the values of parameter $\gamma$, masses $M$, initial coordinate $(t_0, r_0) = (1, 1)$ given in Table \ref{table1} with following eq. (\ref{eq25}).}
	\label{fig4}
\end{figure}
\begin{figure}[h]
	\centering
	\includegraphics{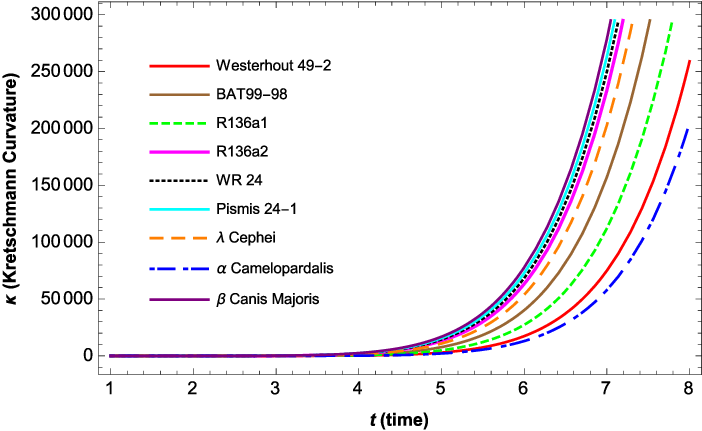}
	\caption{The Kretschmann scalar ($\mathcal{K}$) is plotted with respect to time coordinate \textbf{$t$} for the nine massive stars corresponding to the values of parameter $\gamma$, masses $M$, initial coordinate $(t_0, r_0) = (1, 1)$ given in Table \ref{table1} with following eq. (\ref{eq28}).}
	\label{fig5}
\end{figure}
\begin{figure}[h]
	\centering
	\includegraphics{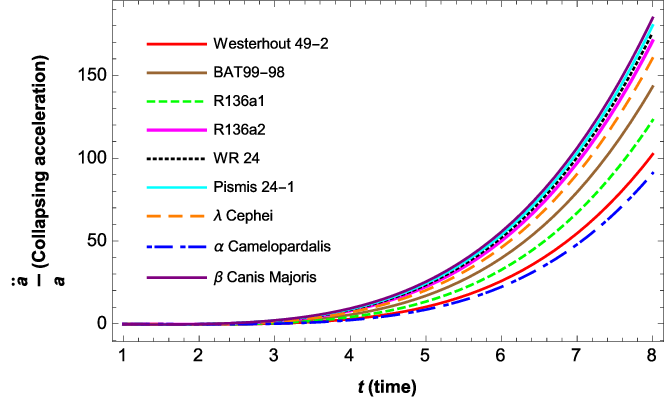}
	\caption{The collapsing acceleration ($\frac{\ddot{a}}{a}$) is plotted with respect to time coordinate \textbf{$t$} for the nine massive stars corresponding to the values of parameter $\gamma$, masses $M$, initial coordinate $(t_0, r_0) = (1, 1)$ given in Table \ref{table1}  with following eq. (\ref{eq91}).}
	\label{fig6}
\end{figure}
\begin{figure}[h]
	\centering
	\includegraphics{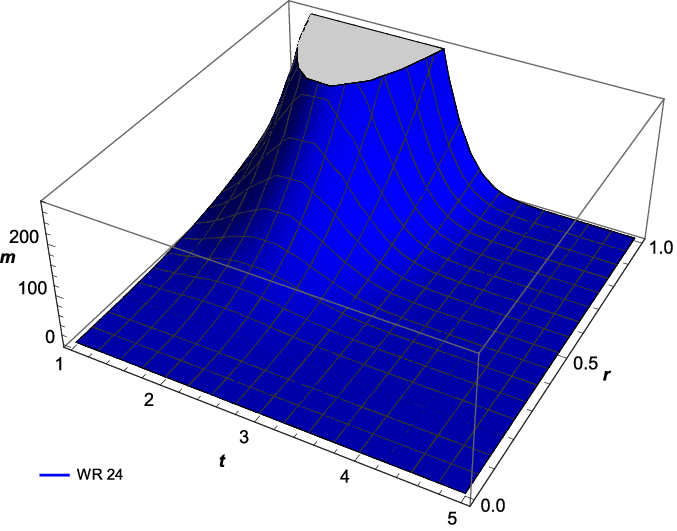}
	\caption{Plot of the mass (\textbf{$m$}) with respect to time coordinate \textbf{$t$} and radial coordinate \textbf{$r$} at center $r = 0$ for the massive stars \textbf{WR 24} corresponding to the values of parameter $\gamma$, mass $M$, initial coordinate $(t_0, r_0) = (1, 1)$ given in Table \ref{table1} with following eq. (\ref{eq27}).}
	\label{fig7}
\end{figure}

\begin{figure}[h]
	\centering
	\includegraphics{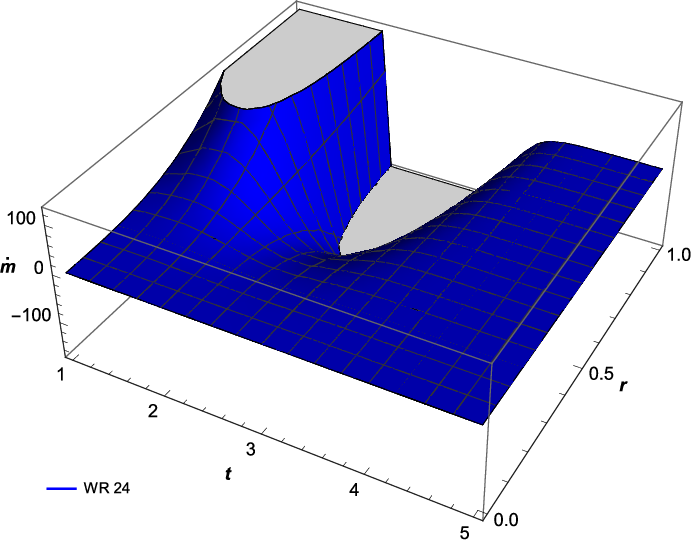}
	\caption{Plot of rate of mass (\textbf{$\dot{m}$}) with respect to time coordinate \textbf{$t$} and radial coordinate \textbf{$r$} at center $r = 0$ for the massive stars \textbf{WR 24} corresponding to the values of parameter $\gamma$, mass $M$, initial coordinate $(t_0, r_0) = (1, 1)$ given in Table \ref{table1} with following eq. (\ref{eq27a}).}
	\label{fig8}
\end{figure}
\begin{figure}[h]
	\centering
	\includegraphics{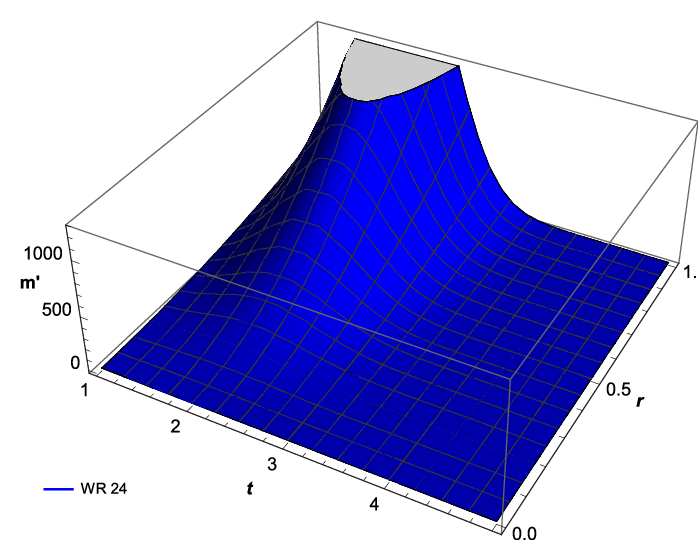}
	\caption{Plot of gradient of mass (\textbf{$m'$}) with respect to time coordinate \textbf{$t$} and radial coordinate \textbf{$r$} at center $r = 0$ for the massive stars \textbf{WR 24} corresponding to the values of parameter $\gamma$, mass $M$, initial coordinate $(t_0, r_0) = (1, 1)$ given in Table \ref{table1} with following eq. (\ref{eq111}).}
		\label{fig9}
	\end{figure}

		\section{Formation of an apparent-horizon and the eternal collapse} \label{sec5}
		\subsection{Apparent horizon}\label{subsec5.1}
		The possible outcomes of gravitational collapse in terms of either a BH or NS are characterized by the occurrence of trapped surface(apparent horizon) developing in the space-time as the collapse progresses. Initially, when object starts collapse under the effect of its own-gravity, no portions of the space-time are trapped but as certain high densities are reached, the trapped surfaces form and a trapped region	develops in the space-time \cite{S2}-\cite{S4}. As it has been seen that the apparent horizon typically develops between the time of  singularity formation and the time at which it meets the outer Schwarzschild event horizon, and the singularity can be either causally connected or disconnected from the outside universe, which is decided by the pattern of trapped surface formation as the collapse evolves \cite{S2}-\cite{S4}.
		\par
		
		In the BH scenario, the apparent horizon forms at a stage earlier than the singularity formation. The outside event horizon then entirely covers the singularity, while the apparent horizon inside the matter evolves from the outer shell to reach the singularity at the instant of its formation \cite{S4}-\cite{S10}. In NS, the trapped surface forms at the centre of the cloud at the time of formation of singularity and the apparent horizon then moves outwards to meet the event-horizon at the boundary in a time later than singularity formation \cite{S10}.
		
		\par
		
		For the space-time metric (\ref{eq1}), the apparent horizon is characterized by\cite{S10}
		\begin{equation}
			R_{,i} R_{,j} g^{ij} = \dot{R}^2(t,r) - 1 = 0
			\label{eq39}
		\end{equation} 
		Since the present study also concerned with the nature of singularity formation due to gravitational collpase of star, we assume that at the initial of the collpase $(t_{0},r_{0})$ the star is not trapped i.e., 
		\begin{equation}
			R_{,i} R_{,j} g^{ij}|_{(t_{0},r_{0})} = r_{0}^2 \dot{a}^2(t_{0}) - 1 < 0
			\label{eq399}
		\end{equation}
		Let us assume that collapsing star forms the apparent horizon surface at $(t_{AH},r_{AH})$, then it follows from  (\ref{eq39}) that 
		\begin{equation}
			\dot{R}^2(t_{AH},r_{AH})-1 = 0
			\label{eq400}
		\end{equation}
		Now using eq.(\ref{eq24}) into (\ref{eq400}), one obtain
		\begin{equation}
			\begin{split}
				\frac{8 \sqrt[3]{\frac{2}{3}} \gamma ^{\frac{2}{3}} M^{\frac{2}{3}} r_{AH}^2 t_{AH}^6 }{3 r_0^2 t_0^4} (8 \gamma  t_0 +5)^{\frac{4}{3}-\frac{125}{1152 \gamma^3}} (8 \gamma  t_{AH}+5)^{\frac{125}{1152 \gamma^3}-2} &\\ e^{ \frac{64 \gamma ^2 t_0^3-60 \gamma  t_0^2+75 t_0+t_{AH} \left(-64 \gamma^2 t_{AH}^2+60 \gamma  t_{AH}-75\right)}{432 \gamma ^2}} = 1		
				\label{eq42}
			\end{split}
		\end{equation}
		The eq.(\ref{eq42}) gives the  finite value of $t_{AH}$, the time formation of  apparent horizon region
		
		\par
		The geometrical radius of apparent horizon surface is 
		\begin{equation}
			\begin{split}		
				R_{AH} = r_{AH} a(t_{AH}) =(\frac{81M}{128 \gamma^2})^{\frac{1}{3}} \frac{r_{AH}}{r_0 t_0^2} (8 \gamma  t_0+5)^{\frac{2}{3}-\frac{125}{2304 \gamma^3}} &\\ e^{\frac{8 \gamma  \left(64 \gamma^2 t_0^3-60 \gamma  t_0^2 + 75 t_0 +t_{AH} (-64 \gamma^2 t_{AH}^2+60 \gamma  t_{AH}-75) \right)+375 log (8 \gamma  t_{AH}+5)}{6912 \gamma^3}}
				\label{eq401}
			\end{split}
		\end{equation}
		The mass-contribution of the collapsing star to the  apparent horizon region is
		\begin{equation}
			\begin{split}
				M_{AH} = m(t_{AH}, r_{AH}) =\frac{M r_{AH}^3 t_{AH}^6}{r_0^3 t_0^6} (8 \gamma  t_0+5)^{2-\frac{125}{768 \gamma^3}} &\\ (8 \gamma  t_{AH}+5)^{\frac{125}{768 \gamma^3}-2} e^{\frac{64 \gamma ^2 t_0^3-60 \gamma  t_0^2+75 t_0+t_{AH} (-64 \gamma^2 t_{AH}^2+60 \gamma  t_{AH}-75)}{288 \gamma^2}}
				\label{eq402}
			\end{split}
		\end{equation}
		It can be seen from eq.(\ref{eq42})  that for the numerical value of model parameter $\gamma$, $t_{AH}$ can be calculated for the massive stars and the radius ($R_{AH}$) and mass ($M_{AH}$) of the apparent horizon region can also be obtained for these stellar system eqs.(\ref{eq401})-(\ref{eq402}).
		
		\subsection{Eternal collapse phenamenon}\label{subsec5.2}
\begin{figure}[h]
			\centering
			\includegraphics[scale=0.6]{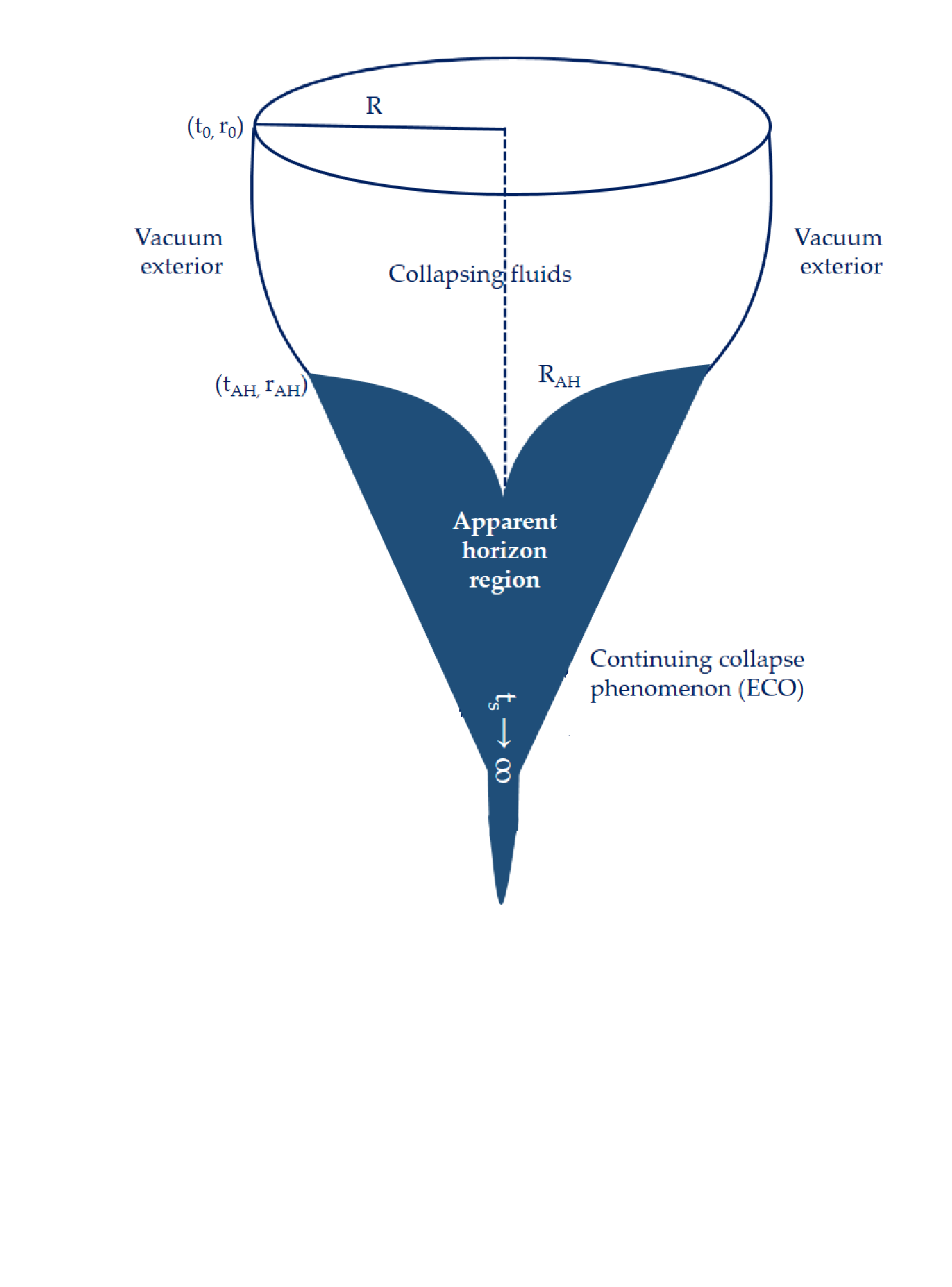}
			\caption{Eternal collapsing object\cite{RJ23a}: The figure shows the continuous collapsing to attain the singularity at infinity}
			\label{fig10}
		\end{figure}

		The gravitational collapse of self-gravitating systems generally end with the formation of a space-time singularity which is
		specified by the divergence of the curvature $\mathcal{K}$ and the energy density $\rho $. Assuming that the star begin to collapse at the moment $t=t_{0}$ where condition (\ref{eq399}) hold i.e., the star is not initially trapped. From eqs.(\ref{eq26}) and (\ref{eq28}) we can see that $\rho \rightarrow \infty $, $\mathcal{K}\rightarrow \infty $ as $t=t_{s}\rightarrow \infty $, in other words the energy density and Kretschmann curvature diverge in  infinite comoving time and hence the star tend to collapse for infinite duration in order to attain the space-time singularity. Further, it can be seen from eq.(\ref{eq42}) that apparent-horizon form in a finite comoving time $t_{AH}$ much earlier than the time of singularity formation ($t_{s}=\infty $). Therefore, the singularity is not naked because before it is formed an apparent-horizon is already formed at $t_{AH}$. Also from (\ref{eq27}), we see that mass vanishes as $t_{s}\rightarrow \infty $. The BH candidate should be formed during gravitational collapse with finite mass in a finite time rather than $m\rightarrow 0$ and $t_{s}\rightarrow \infty $ and therefore, the BH is also not formed here. Also, we see that the acceleration is continuously increases showing accelerating phase of gravitational collapse (as can be seen in figure \ref{fig6}). Hence, one may conclude that homogeneous gravitating star tend to collapse for infinite comoving time in order to attain the singular state $(\rho \rightarrow \infty ,\mathcal{K}\rightarrow \infty )$ and therefore may be called Eternal Collapsing Object (ECO) \cite{RJ23a}\cite{ne34a}. The whole scenario of eternal collapsing star is shown in fig.(\ref{fig10})
		
\section{Discussion and the concluding Remarks}\label{sec6}
The homogeneous gravitational collapse model for the massive stars and its probable end-state are discussed in this work. We have considered the interior of star with the geometry of FLRW space-time and the Schwarzschild geometry is assumed to the vacuum exterior of star. The exact solution is determined by solving the EFEs using the method of parameterization. We have introduced a novel $\Theta$-parameterization as a rational function of $t$ which precisely describes the collapsing process of stellar system (fig.\ref{fig1}).  In addition, we have applied the boundary condition to explicitly obtain the exact solution in terms of mass of star $M$.  In order to assess the astrophysical relevance of our results in terms of graphical depiction, we have taken into consideration the massive stars namely, Westerhout 49-2, BAT99-98, R136a1, R136a2, WR 24, Pismis 24-1, $\lambda$ Cephei, $\alpha$  Camelopardalis, $\beta$ Canis Majoris and throughout the work, all graphs are drawn for these stellar system. Moreover, we have estimated the numerical values of the model parameter $\gamma$ and other variables for these massive stars.
\par
Some of the key features of our model regarding the collapsing configuration are as follows-
\subsection{Graphical aspects}
$\bullet$ The expansion scalar (collapsing rate) are negatively increasing with time during the collapsing process of massive stars (fig.\ref{fig1}). The negative sign of $\Theta$ represents the motion of collapsing fluids towards the centre ($r=0$). \\

$\bullet$ The scale factor ($a$) and areal radius ($R = r a$) are finite within the massive stars and as expected are monotonically decreasing in nature (fig.\ref{fig2}). It can be observe from eq.(\ref{eq24}) that the collapse attains central singularity ($a = 0$) at $t \longrightarrow \infty$.  \\

$\bullet$ From the profile of energy density ($\rho$) and Kretschmann curvature ($\mathcal{K}$) in eqs.(\ref{eq26}) and (\ref{eq28}), we found that both are regular, positive and finite within the stars. Both are increasing in nature and vanish at $t=0$.  It can be observe that $\rho$ and $\mathcal{K}$ both are continuously increasing with time during collapsing configuration and diverge at $t \longrightarrow \infty$ (figs.\ref{fig3} and \ref{fig5}).\\

$\bullet$ The pressure is finite and as expected  it is negatively increasing in nature (fig. \ref{fig4}). The negative sign in eq.(\ref{eq25}) represents the pressure towards the centre($r=0$) during the collapsing configuration.\\

$\bullet$ The acceleration $\frac{\ddot{a}}{a}$ is continuously increasing with time represents the accelerating phase of collapsing configuration (fig.\ref{fig6}). Thus, $\rho, \mathcal{K}$ and $\frac{\ddot{a}}{a}$ are ever-increasing, which tends to extend the physical space-time to an infinite extent, the collapse of sufficiently massive stars may continue forever (fig. \ref{fig10}).\\

$\bullet$ The profile of collapsing mass shows that it is regular, positive and monotonically decreasing with  time and radial coordinate (fig.\ref{fig7}) and $m \longrightarrow 0$ as $t \longrightarrow \infty$. The rate of change of mass  is negative $\dot{m} <0$ and decreases which shows the loss of mass during collapsing configuration(fig.\ref{fig8}). The gradient of mass ($m'$) is also monotonically decreasing(fig.\ref{fig9}). 

\subsection{Dynamics of the model}
The development of the apparent horizon for the scenario of massive stars collapsing has been studied, and its surface radius ($R_{AH}$) and mass ($M_{AH}$) have also been determined eqs.(\ref{eq401})-(\ref{eq402}). It has been demonstrated that the space-time singularity appears as a result of gravitational collapse in an infinite amount of time because the Kretschmann curvature and density have been observed to be divergent at $t=t_s \longrightarrow \infty$.  By comparing the times of singularity formation ($t_s$) with the development of the apparent horizon ($t_{AH}$), we have looked at the final state of the collapsing process in order to draw a definite conclusion concerning singularity formation. By analysis of eq.(\ref{eq42}) one can see that $t_{AH} < t_s$ i.e., the apparent-horizon develops before the singular-state and it turns out into the eternal collapsing process (as discussed in subsec.(\ref{subsec5.2}) and fig. \ref{fig10}). The ECO is the continuing collapse try to attain the singular state in infinite time and its mass $m \longrightarrow 0$ as $t \longrightarrow \infty$.
\par
All of these studies found that our suggested model is highly significant for realistic stellar systems and that it may be employed to explain the phenomena of massive stars collapsing. The importance of our model, however, is that we have provided a fully consistent general relativistic model to describe the collapse scenarios of stellar objects with given masses and radii and, the concept may also be investigated in modified gravitational theories.	
\par
		
\textbf{Acknowledgment:} The authors AJ, RK and SKS are acknowledge to the Council of Science and Technology, UP, India vide letter no. CST/D-2289. Authors RK and SKJP also thankful to IUCAA for all support, where a part of the work done during a visit.   
\par

\textbf{Author contributions}: The authors AJ and RK did all the calculations and graphical representation parts  of article and the draft of the manuscript was written by SKS and SKJ Pacif. All authors read and approved the final manuscript.

\par
\textbf{Funding:} There is no fund available for the publication of this research article.
\par

\textbf{Data Availability Statement:} This manuscript has no associated data or the data will not be deposited.

\par
\textbf{Declarations}
\par
\textbf{Conflict of interest:} The authors have no relevant financial or nonfinancial interests to disclose.

\par
\textbf{Ethical statements:} The submitted work is original and has not been published elsewhere in any form or language (partially or in full).

	\end{document}